\documentclass[aps,psfig,prl,twocolumn,showpacs,groupedaddress]{revtex4}
\usepackage{color} 
\usepackage{xcolor}
\usepackage{amsmath}
\usepackage{graphicx} 
\usepackage{epsfig}

\newcommand{\bee}{\begin{equation}}
\newcommand{\ene}{\end{equation}}
\newcommand{\beea}{\begin{eqnarray}}
\newcommand{\enea}{\end{eqnarray}}

\begin{document}
\title{2-D fluid simulation of a rigid relativistic electron beam driven wakefield in a cold plasma}
\author{Ratan Kumar Bera, Amita Das and Sudip Sengupta}
\affiliation{Institute for Plasma Research, HBNI Bhat , Gandhinagar - 382428, India }
\date{\today}
\begin{abstract} 
Fluid simulations, which are considerably simpler and faster, have been employed to study the behavior of the wake field 
driven by a relativistic rigid beam in a 2-D cold plasma. When the transverse dimensions of the beam is chosen to be  much larger than its longitudinal
extent,  a good agreement with our previous 1-D results  [\textcolor{red}{\it Physics of Plasmas 22, 073109 (2015)}] 
are observed for both under-dense and over-dense beams. 
When the beam is overdense and its transverse extent is smaller or close to the longitudinal
extension, the 2-D blow-out structure, observed in PIC simulations and analytically modeled by Lu et al. [\textcolor{red}{\it Phys. Rev. Lett., 96, 165002 (2006)}]  are recovered. 
For quantitative assessment of particle acceleration in such a wake potential structure test electrons are employed. 
It is shown that the maximum energy gained by the test electrons placed at the back of
the driver beam of energy $\sim 28.5$ GeV, reaches up to $2.6$ GeV in a 10 cm long plasma. These observations are consistent with the experimental results presented in ref. [\textcolor{red}{\it Phys. Rev. Lett. 95, 054802 (2005)}]. 
It is also demonstrated that the  energy gained by the test electrons  get doubled 
($\sim 5.2$ GeV) when the test particles are placed near the axis at the end of the first blowout structure.
\end{abstract}
\maketitle 
\section{ Introduction}
Being an ionized medium, plasma (density  $n_0=10^{18} cm^{-3}$) can support an electric field of the order of hundreds of Giga-Volts in a meter which is several orders of magnitude stronger than that obtained from conventional accelerators \cite{esarey,malkasc}.
This distinct feature of plasma offers a way to design compact and affordable high-performance accelerators in which the charge particles get accelerated by the electric field associated with the plasma wave (so-called ``wakefield''). These accelerating structures or wakefields are excited either using an ultra-intense laser pulse (LWFA) or relativistic electron beam (PWFA) propagating near the speed of light inside the plasma \cite{cj, Golovin,Muggli,Faure,surf}. Here we focus on the excitation of relativistic electron beam driven wakefield, known as plasma wakefield acceleration (PWFA). It was first proposed by Chen, Huff, and Dawson as a means of coupling the relativistic electron beam to the electron plasma wave \cite{chen}. When an electron beam propagates inside a plasma, it expels nearby plasma electrons due to its space charge force. Ions do not respond because of their heavy mass. As the beam (pulsed) passes, the repelled electrons get attracted by the massive ions and hence they come back to neutralize the medium. Further, they overshoot the ions because of their inertia. Finally, these electrons oscillate around their individual ions and sustain a wave in the wake of the beam. Note that the phase velocity of the wake wave is equal to the velocity of the beam \cite{uhm,katsouleas} and independent of plasma parameters (e.g. density and temperature). If an electron 
bunch rides this plasma wave at an appropriate phase, it can be  
accelerated to high energies. This scheme is most suitable to boost the energy of the existing linacs. The success of PWFA scheme has been demonstrated in a number of experiments by accelerating charge particles to GeV energies \cite{hogan,barov,kallos}. The most striking PWFA results were reported by Blumenfeld et al. \cite{Blumenfeld} in 2007, where a 42 GeV electron bunch efficiently accelerates the trailing electrons up to maximum energies of 85 GeV in a meter long plasma. In 2014, Litos et al.\cite{litos} designed an experiment using a separate trailing bunch surfing in an accelerating gradient of 4.4 GV/m. Using the separate bunches, they have minimized the energy spread of the accelerated beam up to $\sim 2 $ percent on average. These results demonstrate the viability of PWFA scheme in high energy physics applications. 
In terms of theoretical work, till date, both linear and nonlinear theory in 1-D has been well established to examine PWFA scheme for several driver configurations over a wide range of beam parameters \cite{rosenzweig,amatuni,ruth,ratan}. Numerical simulations using both fluid \cite{ratan} and Particle-In-Cell (PIC) \cite{epoch} technique for 1-D case have also been carried out. In 2-D, the theoretical studies for the excitation of wakefield have been done only in the linear regime and for specific set of beam configurations \cite{katsouleas_2d,chen_2d}. In the non-linear regime, the exact expression of 2-D wakefields is formidable and hence it has been described by an analytically modeled results given by Lu et al. \cite{lu_prl}. 
However, these theoretical studies on wakefield description have relied on quasistatic approximation where the self-consistent evolution of the beam is ignored. Furthermore,  the non-linear analytical model given by Lu et al. fails to predict the correct form of wakefield structure near the edge of the blow-out.
Therefore, in higher dimensions, the experimental PWFA results are mostly guided by the extensive numerical simulations. Presently, the simulations in this area have been performed using particle-in-Cell (PIC) techniques (e.g. OSIRIS, EPOCH, QUICKPIC etc.) \cite{osiris,epoch,quickpic}. These PIC simulations compute the trajectories of billions of particles for hundreds of plasma periods. Therefore these simulations are computationally heavy and time-consuming. Hence we require powerful computational facilities for such PIC studies. 

\par\vspace{\baselineskip} 
In this paper, we seek the possibility of fluid simulations as simplification in comparison to any sophisticated 
PIC studies, to provide a reasonable empirical guidance on PWFA concept. 
Proposing two-fluid description of plasma wave excitation, two-dimensional fluid simulations have been employed to study the excitation of relativistic electron beam driven wakefield in a cold plasma. We have performed our simulations using a rigid beam for several beam configurations. It is observed that for both under-dense and over-dense beam having a large transverse extension than the longitudinal extension (direction of propagation), the axial profiles of the excited wakefield obtained from our simulation show a good agreement with the 1-D results of Ratan et al. \cite{ratan}. In the other limit i.e. when the transverse dimensions of the beam are smaller or close to its longitudinal extension, the simulation results deviate considerably from the 1-D results. A 2-D analytical study of the linear wake structure is also presented which compares well the simulation results
for small amplitude excitations. It is found that the excitations created by a
short, over-dense beam in simulations are shown to exhibit a  ``blowout'' (a cavity free from cold plasma electrons) structure. 
For over-dense beams, the space-charge force of the beam evacuates all the plasma electrons from its vicinity and creates a pure ion channel behind, which is known as ``blowout''. The key features of the blowout are, i) the mono-energetic electron beam can be self-generated in this region,  ii) the accelerating gradients are nearly constant and the focusing fields are linear inside the blowout, iii) driver beam can propagate many betatron wavelengths without any significant spreading. In recent days,  most of the PWFA experiments are designed and operated in this regime for generating high quality beam \cite{Blumenfeld, litos, hogan}.
We show that the excited blowout wake structure observed in the fluid simulation for a short, overdense beam, also matches with the analytically modeled results of Lu et al. \cite{lu_prl,lu_pop}. These comparisons are made before the wake wave breaks as fluid simulations subsequent to breaking become irrelevant. The breaking of wake wave is identified in the fluid simulation by observing the density profile of plasma which exhibits sharp structures (spiky) just after the breaking\cite{sudip_pre,prabal, prabal_pop_2011,infeld}. In addition, the wave breaking is also observed by tracking the total energy of the system which does not remain conserved after the breaking.  
Further, using the expression for the analytical excursion length of the electrons, we have been able to show that it exceeds the curvature of the blow out structure just before the onset of wave breaking process\cite{bulanov}. In fluid simulations, it is seen that the wake structure or the blow-out structure gets destroyed after the breaking.
However, we have observed that the excitations can survive hundred of plasma periods without any significant deformation for $n_b/n_0 \le 5 $; where $n_b$ and $n_0$ represent the beam density and equilibrium plasma density respectively. 
\par\vspace{\baselineskip} 
Further, we employ test particles (electrons) in the fluid simulation to estimate the energy gain in the process of acceleration.
It is observed that the electrons from the back of the driver beam of energy $28.5$ GeV can gain up to a maximum energy of $2.6$ GeV in a $10$ cm long plasma channel. These results show a good conformity with the experimental results given in ref. \cite{hogan}, 
where an energy gain of $2.6$ GeV in a $10$ cm long plasma is shown. 
We have experimented with the location of injection of the test particles and observe that their energy gain can be doubled to $\sim 5.2$ GeV when they were placed near the axial edge of the first blowout potential structure. 
%
%
The paper has been organized as follows. 
In next section (Section -II), we present the basic governing equations used in the 2-D fluid simulation as well as in the analytical works.  Our numerical observations and associated discussions have been covered in section-III. In section IV, we present the results obtained from the test particle simulations. 
Finally, we have summarized our study in section V.

\par\vspace{\baselineskip}
\section{Governing Equations and Simulation Description}
The basic equations governing the excitation of 2-D relativistic electron beam driven wakefield
in a cold plasma are the relativistic fluid-Maxwell equations. 
The equations contain the equation of continuity and the equation of momentum for plasma electrons.
The dynamics of ion have been ignored because of their heavy mass. They only provide a neutralizing background. In this work, 
the beam has been treated as rigid and its evolution has been ignored. Clearly, this approximation is realistic and 
valid when the beam is  sufficiently energetic  \cite{ratan}. Therefore, the basic normalized governing equations for the excitation of relativistic electron beam driven wakefield in a cold plasma can be written as,    
\begin{equation}
  \frac{\partial n}{\partial t}+\vec{\nabla}. (n\vec{v})=0	
  \label{pl_cont}
 \end{equation}
 \begin{equation}
  \frac{\partial \vec{p} }{\partial t}+(\vec{v}.\vec{\nabla})\vec{p}=-\vec{E}-(\vec{v}\times \vec{B})    
   \label{pl_mom}
 \end{equation}
 \begin{equation}
  \frac{\partial \vec{E} }{\partial t}=(n \vec{v} + n_b \vec{v_b})+(\vec{\nabla}\times \vec{B})    
   \label{elec}
 \end{equation}
  \begin{equation}
  \frac{\partial \vec{B} }{\partial t}=-(\vec{\nabla}\times \vec{E})   
    \label{mag}
 \end{equation}
\begin{equation}
  \vec{\nabla}.\vec{E}=(1-n-n_b)  
     \label{pois}
 \end{equation}
 \begin{equation}
  \vec{\nabla}.\vec{B}=0    
   \label{divb}
 \end{equation}
where $\vec{p}=\gamma \vec{v}$ is the momentum of plasma electron 
having velocity 
$\vec{v}$ and density $n$; where $\gamma=\left(1-v^2\right)^{-1/2}$ is the relativistic factor associated with plasma electron.
Here, $n_b$ and $\vec{v_b}$ represent the density and velocity respectively of the  beam electrons.  
In the above equations, $\vec{E}$ and $\vec{B}$ represent the electric  and magnetic field respectively.
The equations are solved in 2-D geometry (``x-y'' plane) in which the beam propagates along $x$-direction.
The following normalization factors have been used for the above equations,
$t \rightarrow \omega_{pe}t$, $(x,y) \rightarrow \frac{\omega_{pe}(x,y)}{c}$, $\vec{E} 
\rightarrow \frac{e\vec{E}}{m_e c\omega_{pe}}$, $\vec{B} 
\rightarrow \frac{e\vec{B}}{m_e c\omega_{pe}}$,$\vec{v} \rightarrow \frac{\vec{v}}{c}$, 
$\vec{v_b}\rightarrow \frac{\vec{v_b}}{c}$, 
$\vec{p} \rightarrow \frac{\vec{p}}{m_e c}$,$\vec{p_b} \rightarrow \frac{\vec{p_b}}{m_e c}$,
$n\rightarrow \frac{n}{n_0}$, and $n_b\rightarrow \frac{n_b}{n_0}$.
Equations (\ref{pl_cont}-\ref{divb}) are the key equations to study the 2-D excitation 
of relativistic electron beam driven wakefield in a cold plasma. 
\par\vspace{\baselineskip}
The above equations (\ref{pl_cont}-\ref{divb}) are solved using two-dimensional fluid simulation techniques.
The two-dimensional fully electro-magnetic fluid code has been developed using LCPFCT 
suite of subroutines which are based on flux-corrected transport scheme \cite{boris}. The basic principle of this scheme is based on the generalization of two-step Lax-Wendroff method \cite{numr}. The one-dimensional LCPFCT subroutines have been used repetitively to construct the 2-D fluid code by splitting the time steps in the different directions ($x$ and $y$). We have solved the equations (\ref{pl_cont}-\ref{divb})) using this code with non-periodic (open) boundary conditions. Initially, we have introduced the beam at one end of the simulation window which propagates with a constant velocity along $x$-direction towards another end. At each time step, it is checked that the results must satisfy equations (\ref{pois}) and (\ref{divb}). The code has been benchmarked against the widely used 
Particle-In -Cell (PIC) code OSIRIS \cite{osiris} re-producing some standard known results \cite{deepa,deepa2}.   

\par\vspace{\baselineskip}
\section{ Simulation observations and analysis }
In this section, we present our numerical observations and a detail discussion based on these observations. 
The excitation of wakefield is carried out for a bi-Gaussian beam having density profile  $n_b=n_{b0} exp(-\frac{x^2}{2\sigma_x^2}) exp(-\frac{y^2}{2\sigma_y^2})$; where $n_{b0}$, $\sigma_x$ and $\sigma_y$ represent the peak value of the beam density, longitudinal width of beam and transverse width of the beam respectively. Below we present our simulation results studying different aspects on relativistic electron beam driven wakefield excitations. 
\subsection{Effect of finite transverse beam size}
\begin{figure*}
    \includegraphics[width=0.9\textwidth]{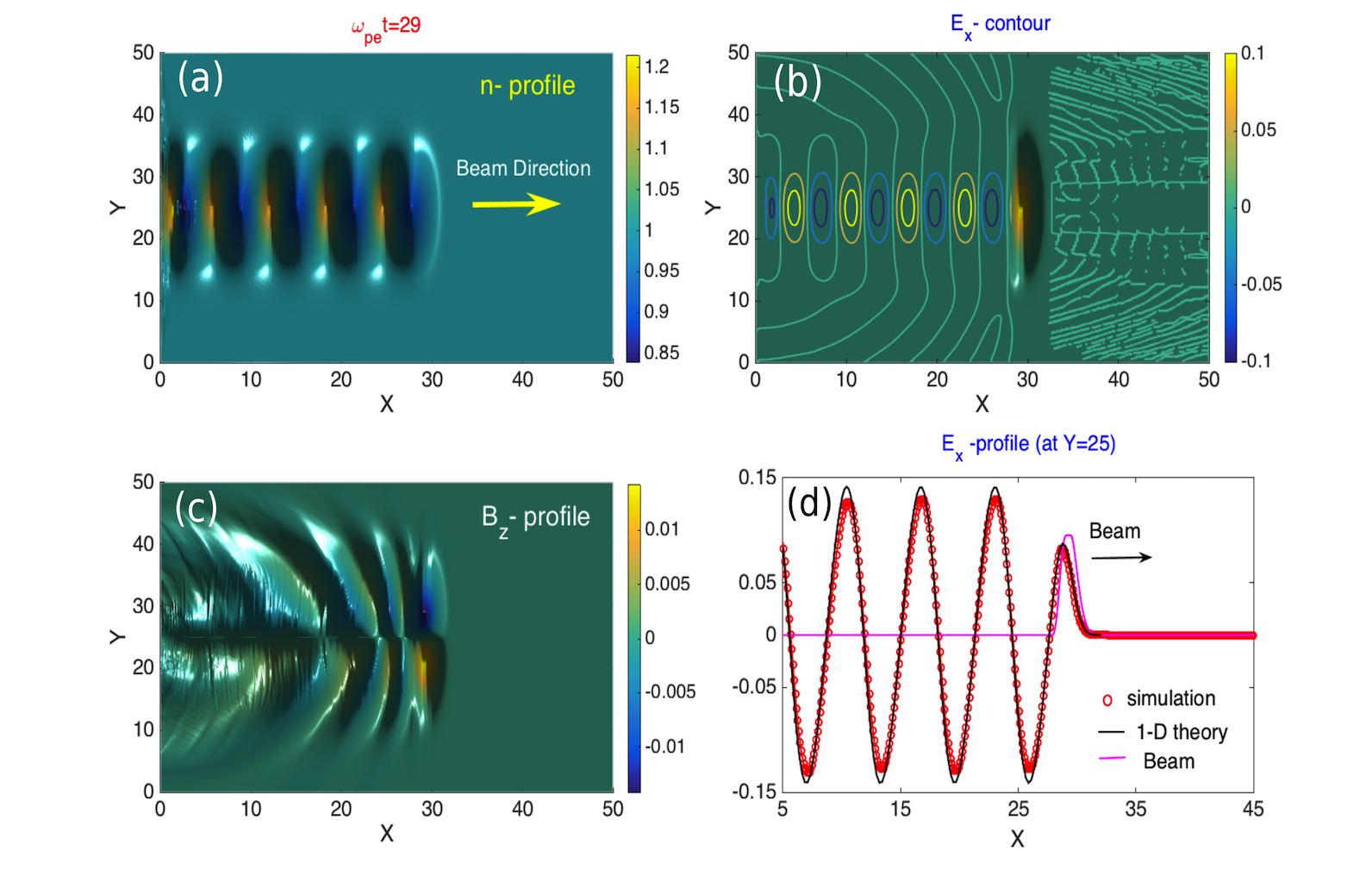}
    \caption{Plot of (a) normalized electron density ($n$), (b) longitudinal electric field ($E_x$), (c) $z-$ component of magnetic field ($B_z$), and (d) axial profiles (integrated in $Y$-direction) of longitudinal electric field profile at $t=29$ for a bi-Gaussian beam of normalized peak density ($n_{b0}$)=0.1, beam velocity ($v_b$) =0.9999, $\sigma_x=0.5$ and $\sigma_y=\sqrt{15}$.}
    \label{fig1}
\end{figure*}

\begin{figure*}
    \includegraphics[width=0.9\textwidth]{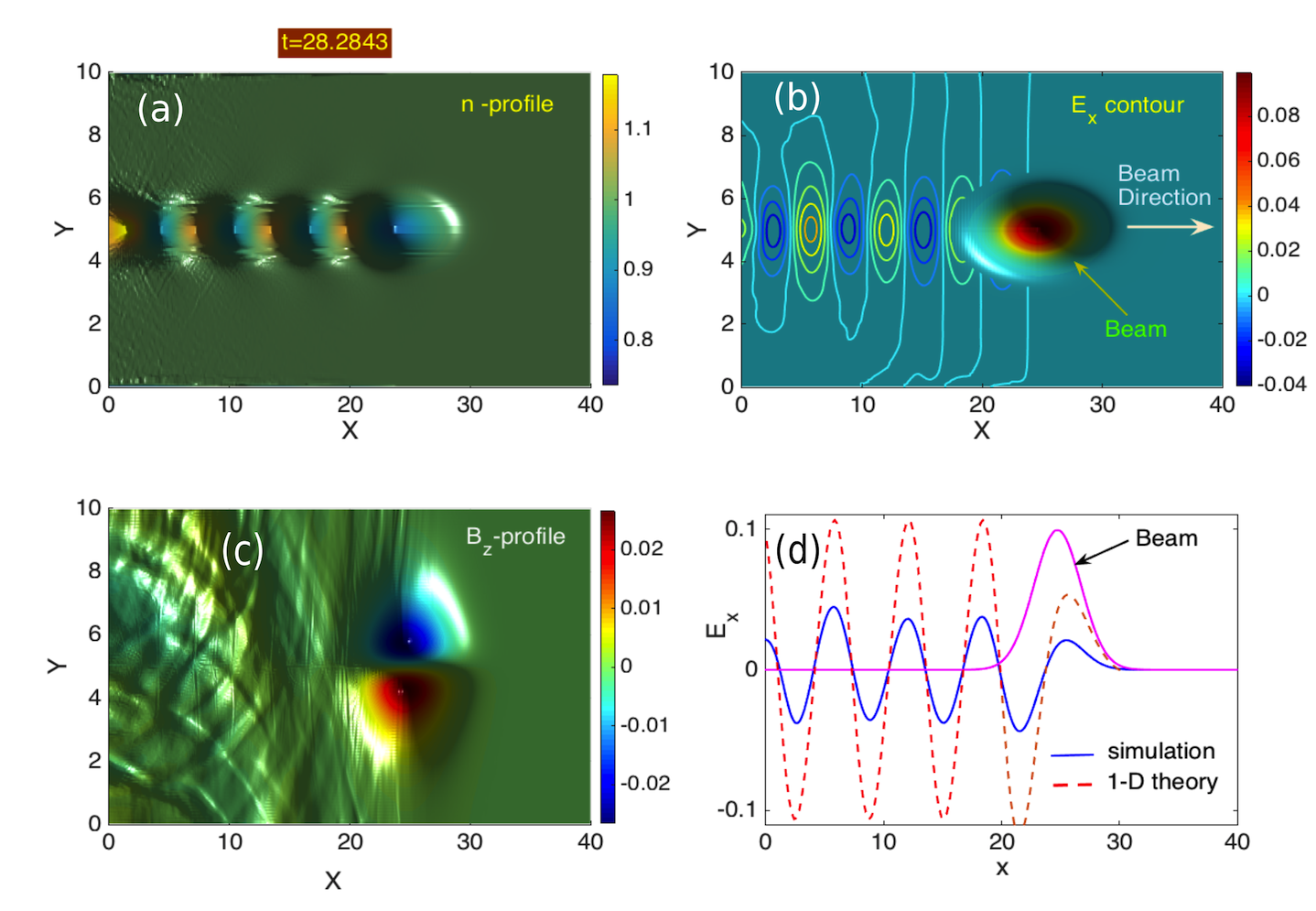}
    \caption{Plot of (a) normalized electron density ($n$), (b) longitudinal electric field ($E_x$), (c) $z-$ component of magnetic field ($B_z$), and (d) axial profiles (integrated in $Y$-direction) of longitudinal electric field profile at $t=28.28$ for a bi-Gaussian beam of normalized peak density ($n_{b0}$)=0.1, beam velocity ($v_b$) =0.9999, $\sigma_x=\sqrt{5}$ and $\sigma_y=0.5$.}
    \label{fig2}
\end{figure*}

First, we have performed our simulations for different beam length ratios which is defined as, $l_s=\frac{\sigma_x}{\sigma_y}$.
Figs. (\ref{fig1}) and (\ref{fig2}) shows the excited wakefield profiles in terms of electron density ($n$), longitudinal electric field ($E_x$), and $z$-component of magnetic field ($B_z$) for $ l_s=\frac{0.5}{\sqrt{15}}=0.129 < 1$ and $l_s=\frac{\sqrt{5}}{0.5}=4.4 >1$ respectively; where the peak density of the beam $n_{b0}=0.1$  and velocity  $v_b=0.9999$ in both the cases. The last sub-plot (d) of these Figs. (\ref{fig1}) and (\ref{fig2}) shows the axial profile of longitudinal electric field ($E_x$) obtained from our 2-D simulation by integrating along the transverse directions and the 1-D profile given in ref. \cite{ratan} for the same values of the beam parameters. It is seen that the simulation results match with the 1-D results for $l_s<1$ and deviate for $l_s>1$. It indicates that, for a beam having a small transverse extension compared to its longitudinal length, the effect of transverse magnetic field plays an important role. The transverse magnetic field bends the trajectories of the electrons. Hence the charge separation decreases in the longitudinal direction and we observe a significant suppression in the amplitude of the longitudinal electric field in 2-D than that obtained from 1-D theory.  
\par
\begin{figure*}
    \includegraphics[width=0.9\textwidth]{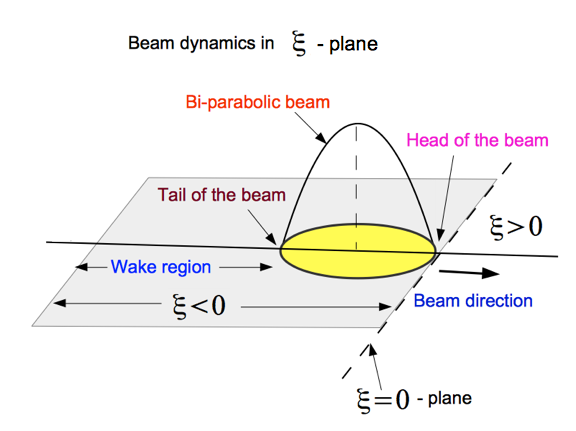}
    \caption{Schematic diagram of beam dynamics in ($\xi$, $r$)-frame.}
    \label{fig3}
\end{figure*}

\vspace{\baselineskip}
For the sake of completeness, an analytical solution of relativistic electron beam driven wakefield in the linear regime (i.e. $n_b<<1$) is also presented in this paper. The analytical solutions are obtained in cylindrical co-ordinates ($r$, $\theta$, $x$) system corresponding to cartesian co-ordinates ($x$, $y$, $z$); where $r=\sqrt{y^2 +z^2}$ and $\theta=tan^{-1}(\frac{z}{y})$ represent the radial and azimuthal coordinates respectively. We have also considered the azimuthal ($\theta$) symmetry in the following derivation i.e. $\partial/ \partial \theta \equiv 0$. 
Following the method given in ref. \cite{chen_2d}, the equation of continuity for perturbed plasma density ($n^{in}_1=n^{in}-1$) inside the beam in a co-moving frame ( $\xi=x-v_bt$, $r$) can be written as,
\begin{equation}
\partial^2_\xi n^{in}_1(\xi,r) + n^{in}_1(\xi, r)=-n_b(\xi,r)        \hspace{1cm} -\xi_f \le \xi \le 0 
\label{n_1}
\end{equation}
where $v_b=1$ and $\xi_f=2\pi \omega_{pe} l_b/c$ defines the value of $\xi$ at the tail of the beam having physical length $l_b$. A schematic diagram of the dynamics of the beam in ($\xi$, $r$) frame is shown in Fig. (\ref{fig3}). In this frame, the beam exists in the region $ -\xi_f \le \xi \le 0$. As the beam propagates to a speed equal to the speed of light, the profile of $n_1 (\xi,r)=0$, $\vec{v}(\xi,r)=0$, $\vec{E}(\xi,r)=0$ and $\vec{B(\xi,r)}=0$, at the front of the beam ($\xi>0)$. 
In this frame ($\xi$, $r$), using the relations $\vec{B}=\vec{\nabla} \times \vec{A}$ and $\vec{E}=-\vec{\nabla} \phi -\frac{\partial \vec{A}}{\partial t}$, the Eqs. (\ref{elec}-\ref{divb}) inside the beam( $-\xi_f \le \xi \le 0$) can be reduced to the following form.
\begin{equation}
(\nabla^2_r -1)(A^{in}_{1x}-\phi^{in}_1)=-n^{in}_1      \hspace{1cm} -\xi_f \le \xi \le 0 
\label{lfield}
\end{equation}
where $\phi^{in}_1$ and $A^{in}_{1x}$ represent the perturbed values of scalar potential ($\phi$) and $x$-component of vector potential ($\vec{A}$) inside the beam respectively.  
The solution of the above Eq. (\ref{n_1}) for a given $n_b$ provides the form of perturbed plasma density $n^{in}_1 (\xi,r)$ inside the beam $ -\xi_f \le \xi \le 0$. Substituting the form of $n^{in}_1$ in the R.H.S of the Eq. (\ref{lfield}), the solution of ($A^{in}_{1x}-\phi^{in}_1$) can be obtained inside the beam. The longitudinal electric field ($E^{in}_{1x}$) inside the beam can be obtained from the following equation.
\begin{equation}
E^{in}_{1x}=\frac{\partial}{\partial \xi} (A^{in}_{1x}-\phi^{in}_1)
\label{gelec}
\end{equation}

\par\vspace{\baselineskip}

At the wake of the beam i.e. in the region $-\infty \le \xi \le -\xi_f$, the density of the beam $n_b=0$. Thus the equations (\ref{n_1}) and (\ref{lfield}) outside the beam $-\infty \le \xi \le -\xi_f$  can be written as,
\begin{equation}
\partial^2_\xi n^{wake}_1(\xi,r) + n^{wake}_1(\xi, r)=0        \hspace{1cm} -\infty \le \xi \le -\xi_f
\label{n_1w}
\end{equation}
\begin{equation}
(\nabla^2_r -1)(A^{wake}_{1x}-\phi^{wake}_1)=-n^{wake}_1      \hspace{1cm} -\infty \le \xi \le -\xi_f 
\label{wlfield}
\end{equation}
 where $n^{wake}_1$, $\phi^{wake}_1$ and $A^{wake}_{1x}$ represent the perturbed values of plasma density ($n$), scalar potential ($\phi$) and $x$-component of vector potential ($\vec{A}$) at the wake of the beam respectively. Replacing the form of $(A^{in}_{1x}-\phi^{in}_1)$  with $(A^{wake}_{1x}-\phi^{wake}_1)$ in Eq. (\ref{gelec}), the expression of longitudinal electric field ($E^{wake}_{1x}$) can be obtained outside the beam.
 
\par\vspace{\baselineskip}
The exact analytical solution is obtained here for a bi-parabolic beam having density profile $n_b=n_{b0}(1-\frac{(\xi+b)^2}{b^2})(1-\frac{r^2}{a^2}) = n_{b0} g(\xi) f(r)$; where $a$ and $b$ defines the extension of the beam along $\xi$ and $r$ respectively. 
Using Eq.(\ref{n_1}), the form of $n^{in}_1$ inside the beam can be written in terms of Green function as, 
\begin{equation}
n^{in}_1= -n_{b0} f(r) \int_{\xi} ^{\infty}  d\xi' g(\xi')sin (\xi'-\xi) =-n_{b0} f(r)G(\xi) ;  \hspace{1cm} -2b \le \xi \le 0 
\end{equation}
At the front of the beam ($\xi>0$),  the perturbed plasma density $n_1(\xi, r)=0$, as the perturbation can not travel faster than speed of light. The solution of $G(\xi)$ for a bi-parabolic can be written as,

$$G(\xi)=\int_{\xi} ^{0} \left(1-\frac{(\xi'+b)^2}{b^2}\right) sin (\xi'-\xi) d\xi' $$
$$=1-\frac{(\xi+b)^2}{b^2}+\frac{2}{b} sin (\xi) + \frac{2}{b^2}(1-cos(\xi))$$

Therefore, the form of perturbed plasma density inside the beam is,
\begin{equation}
\begin{matrix}
n^{in}_1= -n_{b0} \left(1-\frac{r^2}{a^2}\right)[1-\frac{(\xi+b)^2}{b^2})+\frac{2}{b} sin (\xi) \\
 + \frac{2}{b^2}(1-cos(\xi)] ; 
 \hspace{1cm}  -2b \le \xi \le 0 
 \end{matrix}
 \label{ninpara}
\end{equation}
Substituting the form of $n_1^{in}$ in Eq. (\ref{lfield}), the solution of $(A^{in}_{1x}-\phi^{in}_1)$  can be written as,
\begin{equation}
(A^{in}_{1x}-\phi^{in}_1)=-n_{b0} G(\xi) F(r)  \hspace{1cm} -2b \le \xi \le 0 
\label{ain}
\end{equation}
Where $F(r)=\int_{0} ^{r} r'f(r') I_0 (r')K_0(r) dr' +  \int_{r} ^{\infty} r'f(r') I_0 (r)K_0(r') dr'$. Here $I_0$ and $K_0$ represent modiefied  Bessel function of first and second kind respectively. 
Integrating the equation of $F(r)$ in the range $0<r<a$, we get,
$$F(r)=2\left[I_0(r)K_0(a) +\frac{1}{2}(1-\frac{r^2}{a^2}) -\frac{2}{a^2}\right]$$
Substituting the form of $F(r)$ in Eq.(\ref{ain}), we have,

\begin{equation}
\begin{matrix}
(A^{in}_{1x}-\phi^{in}_1)=-2n_{b0} \left((1-\frac{(\xi+b)^2}{b^2})+\frac{2}{b} sin (\xi) + \frac{2}{b^2}(1-cos(\xi)\right) \\
\times \left[I_0(r)K_0(a) +\frac{1}{2}(1-\frac{r^2}{a^2}) -\frac{2}{a^2}\right]  \hspace{1cm} -2b \le \xi \le 0 
\end{matrix}
\label{apain}
\end{equation}

\begin{figure*}
    \includegraphics[width=0.9\textwidth]{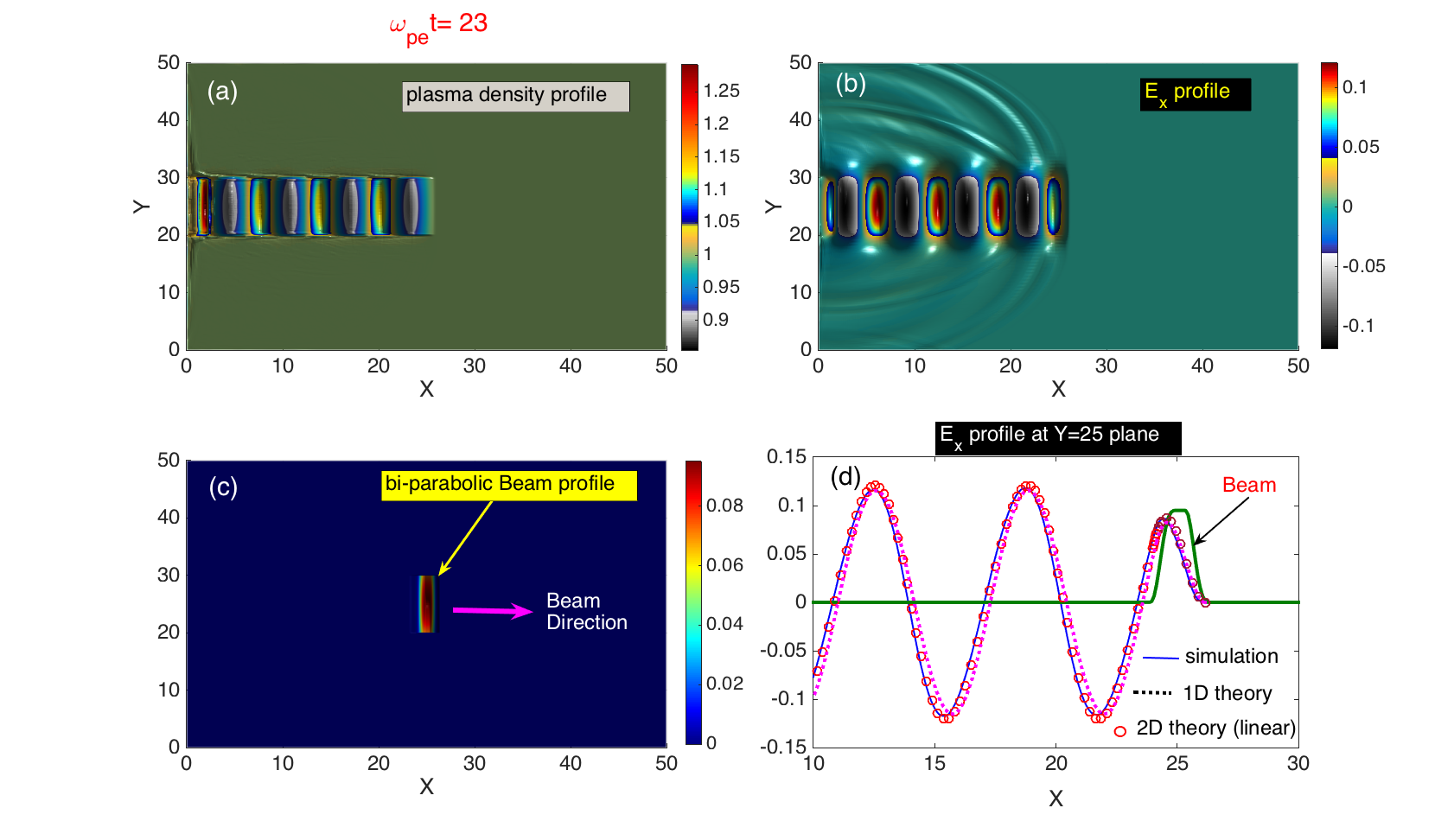}
   \caption{Plot of (a) normalized electron density ($n$), (b) longitudinal electric field ($E_x$), (c) $z-$ component of magnetic field ($B_z$) profile and (d) axial profiles (integrated in $Y$-direction) of longitudinal electric field profile at $t=23$ for a bi-parabolic beam of normalized peak density ($n_{b0}$)=0.1, beam velocity ($v_b$) =0.9999, $b=0.5$ and $a=\sqrt{15}$.}
    \label{fig4}
\end{figure*}

\begin{figure*}
    \includegraphics[width=0.9\textwidth]{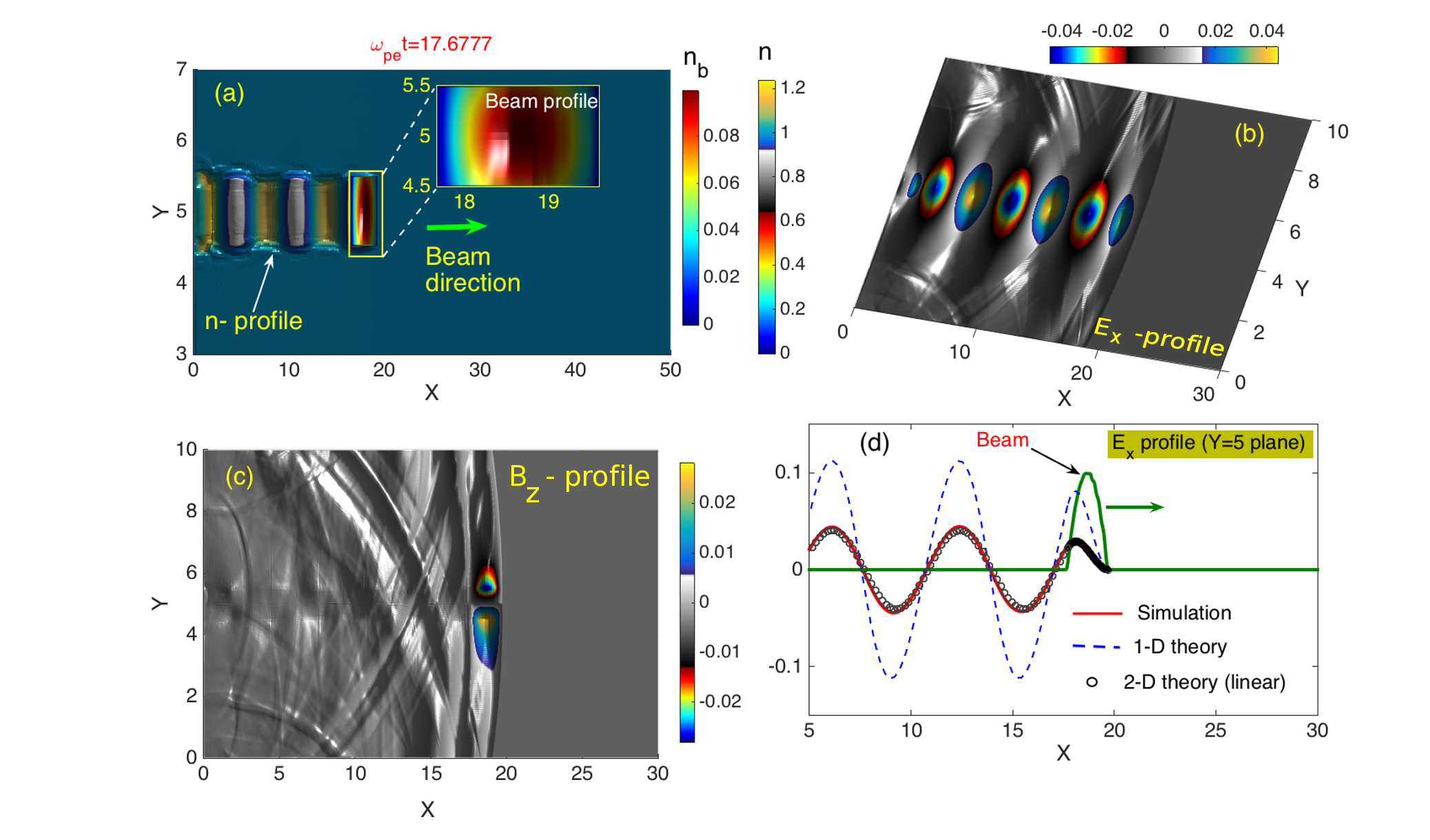}
  \caption{Plot of (a) normalized electron density ($n$), (b) longitudinal electric field ($E_x$), (c) $z-$ component of magnetic field ($B_z$), and (d) axial profiles (integrated in $Y$-direction) of longitudinal electric field profile at $t=17.67$ for a bi-parabolic beam of normalized peak density ($n_{b0}$)=0.1, beam velocity ($v_b$) =0.9999, $b=\sqrt{5}$ and $a=0.5$.}
    \label{fig5}
\end{figure*}

The longitudinal electric field inside the beam is,
 \begin{equation}
 \begin{matrix}
E^{in}_{1x}=\frac{\partial}{\partial \xi}(A^{in}_{1x}-\phi^{in}_1) 
=2n_{b0} \left[I_0(y)K_0(a) +\frac{1}{2}(1-\frac{y^2}{a^2}) -\frac{2}{a^2}\right] \\
\times\left[(-\frac{2(\xi+b)}{b^2})+\frac{2}{b} cos (\xi) + \frac{2}{b^2}sin(\xi)\right] ; \hspace{0.5cm} -2b \le \xi \le 0 
\end{matrix}
\label{elecinside}
\end{equation}
At the wake of the beam ($n_b=0$), the solution of the equation (\ref{n_1w}) can be written as, 
\begin{equation}
n^{wake}_1=A(r) sin (\xi) +B(r) cos (\xi) ;     \hspace{0.5cm} -\infty \le \xi \le -2b 
\label{wake_n}
\end{equation}

where $A(r)$ and $B(r)$ are the integration constants. It is to be noted that the perturbed plasma density ($n_1$) and the derivative of the plasma density w.r.t. $\xi$  has to be continuous at the end of the beam $\xi=-2b$. Therefore, we have $n^{wake}_1(-2b, r)=n^{in}_1(-2b, r)$ and $\partial_\xi n^{wake}_1(-2b, r)=\partial_\xi n^{in}_1(-2b, r)$. Using these conditions, we get, 

$$A(r) sin(-2b)+B(r) cos(-2b)=-n_{b0} f(r) G(-2b)$$
$$A(r) cos(-2b)-B(r) sin(-2b)=-n_{b0} f(r) G'(-2b)$$
Here `prime' represents the differentiation w.r.t. $\xi$. Solving the above equations of $A(r)$ and $B(r)$, we get, 
$A(r)=-n_{b0}f(r) (G(-2b)sin (-2b) + G'(\xi_f) cos (-2b))$ and $B(r)=-n_{b0}f(r) \left(G(-2b) cos (-2b) -sin(-2b)G'(-2b)\right)$.
Substitutuing the form of $n_1$ and integratiing the equation (\ref{wlfield}), we get,
\begin{equation}
A^{wake}_{1x}-\phi^{wake}_1=M(r) sin (\xi) +N(r) cos (\xi) ;    \hspace{0.2cm} -\infty \le \xi \le -2b
\end{equation}
Where, 
\begin{equation}
M(r)=-n_{b0}F(r) (G(-2b)sin (-2b) + G'(-2b) cos (-2b))
\end{equation}
\begin{equation}
N(r)=-n_{b0}F(r) \left(G(-2b) cos (-2b) -sin(-2b)G'(-2b)\right)
\end{equation}
Thus the longitudinal electric field at the wake of the beam can be derived as,
\begin{equation}
\begin{matrix}
E^{wake}_ {1x}(\xi,r)=M(r) cos (\xi) - N(r) sin (\xi)
\end{matrix}
\label{wake_E}
\end{equation}
Using a bi-parabolic beam, we have performed our fluid simulations for different beam length ratios. In simulations, the observations are made in ($x$, $y$)-plane  ( i.e. $\theta=0$ plane in cylindrical co-ordinates).
The simulation results are shown in Figs. (\ref{fig4}) and (\ref{fig5}) for  $ l_s=\frac{b}{a}=0.129 <1$ and $l_s=\frac{b}{a}=4.4 >1$ respectively; where $n_{b0}=0.1$ and $v_b=0.9999$ in both the cases.
In the last subplots (d) of Figs.(\ref{fig3}) and (\ref{fig4}), we have plotted the axial profile ($y=0$ plane) of longitudinal electric field obtained from our simulation along with the 2-D linear analytical profile obtained from Eqs. (\ref{elecinside}) and (\ref{wake_E}) at $y=0 (r=0)$ on the top of 1-D theoretical profile given in ref. \cite{ratan}. The 1-D results are obtained by solving Eqs. (8) in ref.\cite{ratan} both inside and outside the beam for a bi-parabolic beam.
We have observed that the simulation result matches with the 2-D theoretical results for any arbitrary values of $b/a$ in the linear regime. This validates our simulation results for the excitation of relativistic electron beam driven wakefield in a cold plasma. For a beam having  $b/a \ge1$ (see Fig. \ref{fig5} (d)), the longitudinal electric field profile obtained from the simulation deviates from 1-D result. This certainly concludes that the finite transverse size of the beam plays an important role in the excitation. For a beam having transverse size larger than the longitudinal extension, the excitation tends to be more electrostatic in nature.  On the other hand, the excitations exhibit electromagnetic nature for a beam having transverse size smaller or equal to the longitudinal extension. This happens due to finite transverse size of the beam for which the electric fields acquires a curvature leading to the appearance of the transverse magnetic fields. These transverse magnetic fields restrict the longitudinal movement of electrons. Hence the charge separation in longitudinal direction decreases. Therefore the amplitude of the longitudinal electric field in 2-D decreases significantly than that obtained from 1-D cases (see Fig. \ref{fig5} (d)).

\par\vspace{\baselineskip}
\subsection{Fluid simulation in the blowout regime}
\begin{figure*}
    \includegraphics[width=0.9\textwidth]{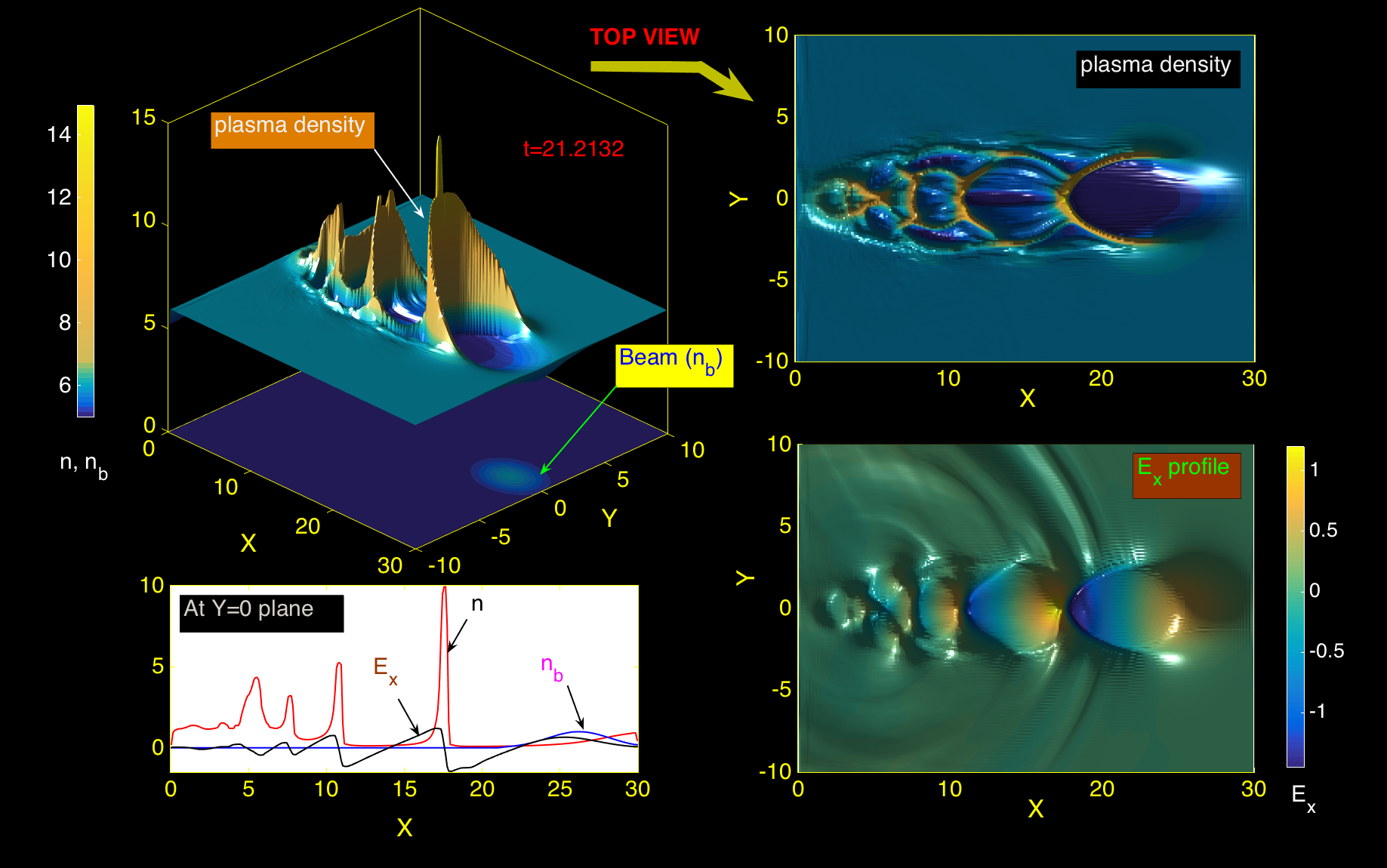}
    \caption{Plot of normalized perturbed plasma electron density ($n$) profile, longitudinal electric field ($E_x$) profile  for a bi-Gaussian beam of normalized peak density ($n_{b0}$)=2, beam velocity ($v_b$) =0.9999, $\sigma_x=1$ and $\sigma_y=0.4$.}
    \label{fig6}
\end{figure*}

\begin{figure*}
    \includegraphics[width=0.9\textwidth]{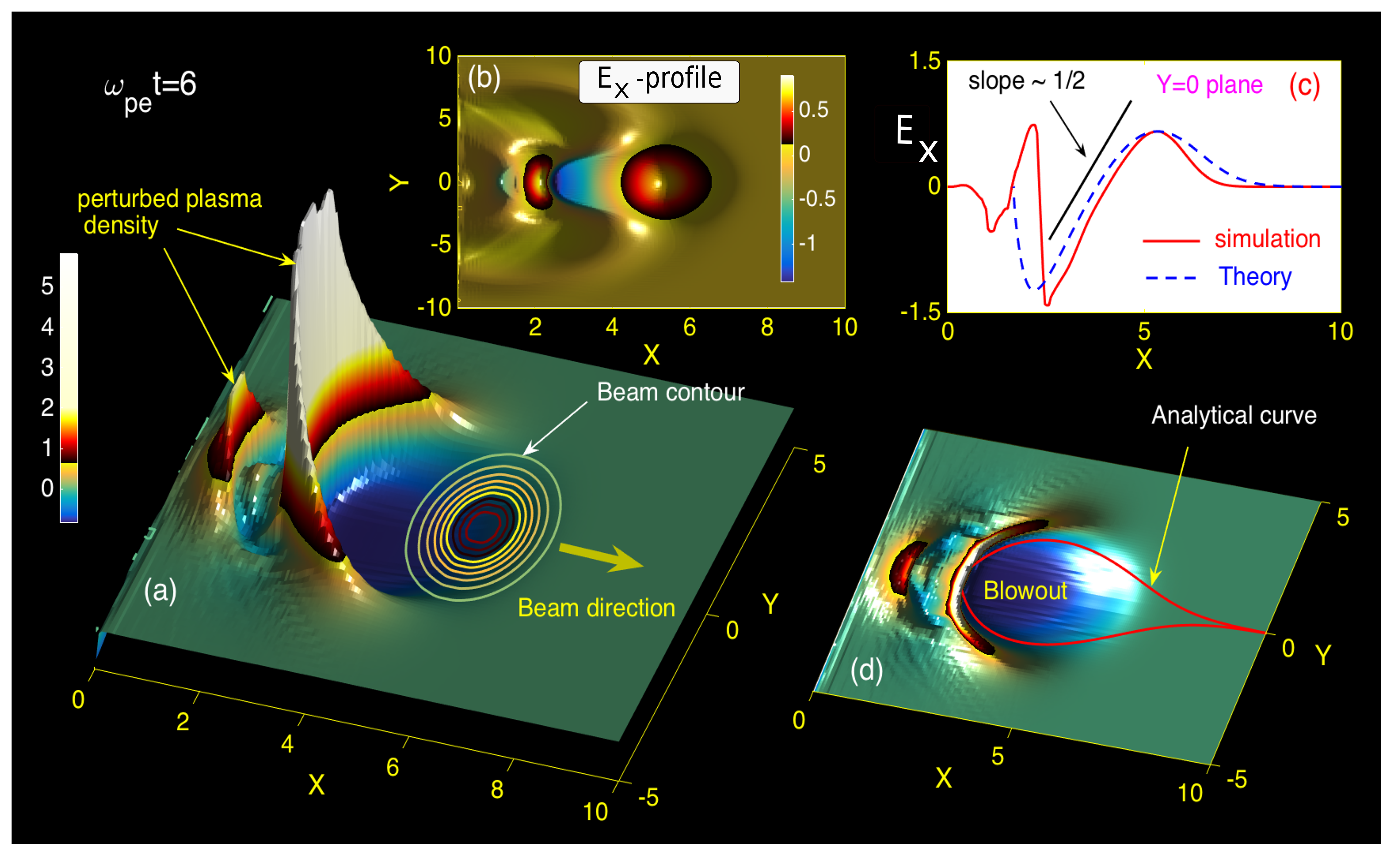}
    \caption{Plot of  (a) normalized perturbed electron density ($n_1$) profile, (b) longitudinal electric field profile, (c) axial profile of longitudinal electric field profile, and (d) analyticall obtained blowout curve (red line) at $t=6$ for a bi-Gaussian beam of peak density ($n_{b0}$)=1, beam velocity ($v_b$) =0.9999, $\sigma_x=\sqrt{2}$ and $\sigma_y=1$.}
    \label{fig7}
\end{figure*}

\begin{figure*}
    \includegraphics[width=1.0\textwidth]{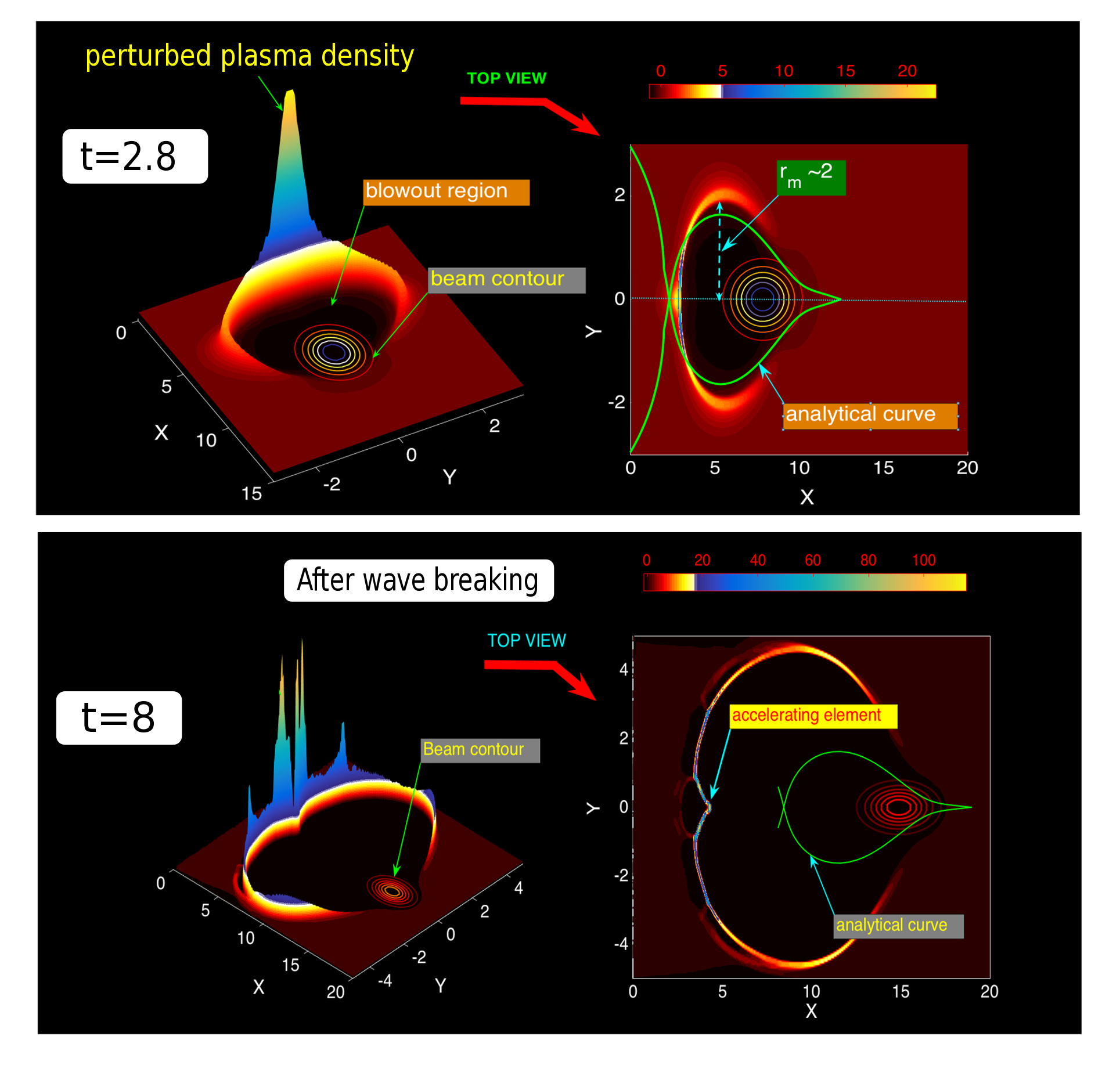}
    \caption{Plot of normalized perturbed electron density ($n_1$) profile at different times for the normalized beam density ($n_b$)=7.0, $\sigma_x=\sqrt{2}$ and $\sigma_y= 0.4$.}
    \label{fig8}
\end{figure*}

\begin{figure*}
    \includegraphics[width=0.9\textwidth]{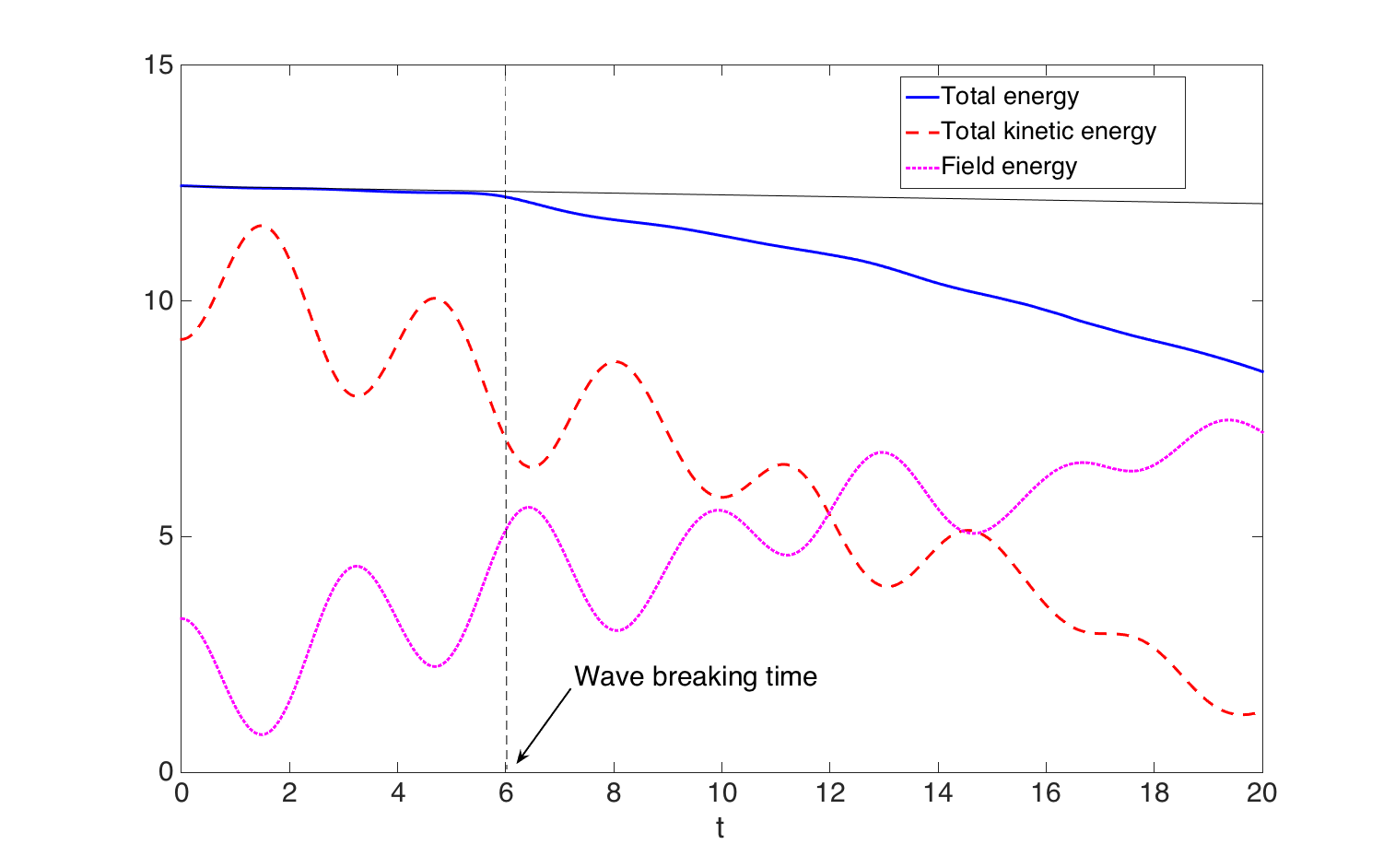}
   \caption{Plot of normalized total energy, kinetic energy and field energy as a function of time for the normalized beam density ($n_b$)=7.0, $\sigma_x=\sqrt{2}$ and $\sigma_y=0.4$.}
    \label{fig9}
\end{figure*}

Here we present the excitation of wakefield driven by a relativistic electron beam having density higher than the equilibrium plasma density.
In all cases, the fluid simulations are performed using a rigid, bi-gaussian beam. In Fig. (\ref{fig6}), we plot the excitation of wakefield in terms of longitudinal electric field and perturbed plasma density profile for $n_{b0}=2$, $v_b=0.9999$, $\sigma_x=1$ and $\sigma_y=0.4$. It is observed that the excitation exhibits blowout structure in the simulation.
Fig. (\ref{fig7})  shows the excitation for  $n_{b0}=1$, $v_b=0.9999$, $\sigma_x=\sqrt{2}$ and $\sigma_y=0.4$. We have also plotted the analytical profile of longitudinal electric field (dotted blue line) in  Fig. (\ref{fig7}(c)) and a curve of blowout (solid red line) in (\ref{fig7}(d)) obtained from the analytical modeled results presented in ref. \cite{lu_prl,lu_pop}. The profile of analytical blowout curve and longitudinal electric field is obtained by solving the equations (46) and (47) given in ref. \cite{lu_pop}, using the same parameter values. It is seen that our simulation results show a good agreement with the analytical results. However, for a beam density $n_{b0}=7$, $v_b=0.9999$, $\sigma_x=\sqrt{2}$ and $\sigma_y=0.4$,  the numerical results deviate from analytical theory after several plasma periods (see Fig. \ref{fig8}). For a sufficiently intense beam, the beam expels all nearby plasma electrons which get accumulated near the edge of the blowout.  The size of the blowout gradually increases and eventually gets destroyed exhibiting sharp spikes in the density profile.This is a clear signature of wave breaking \cite{sudip_pre,prabal, prabal_pop_2011,infeld}. 
Recently, Deepa et al. \cite{deepa2} have shown that the total energy of the system must drop down after the wave breaking in the fluid simulation. This happens because wave breaking results in transfer of energy to high wave number which can not be resolved by the chosen grid size. The fluid description is no longer valid after the wave breaking. The total energy of the system ($T_{tot}$) in our simulation at a given time $t$ can be written as,
\begin{equation}
T_{tot}= T_{kin}+ T_{field}
\label{TE}
\end{equation}
where $T_{kin}$ and $T_{field}$ are the kinetic and field energy of the system respectively. The kinetic energy and the filed energy of the system is given by,
\begin{equation}
\begin{matrix}
T_{kin}=\sum\limits_{i=1}^{N_x} \sum\limits_{j=1}^{N_y} \left[n(i,j)(\gamma(i,j)-1)\right] \Delta x \Delta y \\ + \sum\limits_{i=1}^{N_x} \sum\limits_{j=1}^{N_y} \left[n_b(i,j)(\gamma_b(i,j)-1)\right] \Delta x \Delta y
\end{matrix}
\label{KE}
\end{equation}
 \begin{equation}
T_{field}=\sum\limits_{i=1}^{N_x} \sum\limits_{j=1}^{N_y} \left[\frac{E^2(i,j) +B^2(i,j)}{2}\right] \Delta x \Delta y 
\label{FE}
\end{equation}
where $E^2=E_x^2 +E_y^2 + E_z^2$ and  $B^2=B_x^2 +B_y^2 + B_z^2$ are the square of the magnitude of the electric and magnetic field respectively. Here $i$ and $j$ represent the index (integers) corresponding to the grid numbers along $x$ and $y$-directions respectively. $N_x$ and $N_y$ are the total number of grid points along $x$ and $y$- directions in the system respectively. The second term in the R.H.S of the Eq. (\ref{KE}) represents the energy term for the driver beam.
At each time, we have numerically calculated the total energy using the above expressions of energy term for beam density $n_{b0}=7$, $v_b=0.9999$, $\sigma_x=\sqrt{2}$, $\sigma_y=0.4$ and plotted in Fig. (\ref{fig9}) as a function of time ($t$). 
It is observed that the total energy of the system drops down just after the wave breaking as shown in Fig. (\ref{fig9}).
Further, the wake wave breaking time can be obtained by calculating the excursion length of the electrons.  It is well known that the wake wave breaks when the excursion length exceeds the value of the radius of curvature  \cite{bulanov}.  In Fig. (\ref{fig8}), it is clear that the wave first breaks at the axial edge of the blowout. By fitting a circle at the axial edge of the blowout before it gets destroyed, we have calculated the value of the radius of curvature $R=3.5$. The excursion length is defined as  $l_e=\frac{E_{max}}{\omega^2}$; where $E_{max} $ and $\omega$ are the maximum electric field at the blowout and the characteristic frequency of the wake wave respectively. Before the wave breaking time i.e. at $t=2.8$, we have found $l_e=2.9<R$, and $l_e=4.2>R$ after the wave breaking ($t=8$). However, for a beam density $n_{b0} \le 5$, it is observed that the blowout can survive hundreds of plasma periods without any significant deformation in the fluid simulation.

\section{Test particle simulation and results}
\begin{figure*}
    \includegraphics[width=1.0\textwidth]{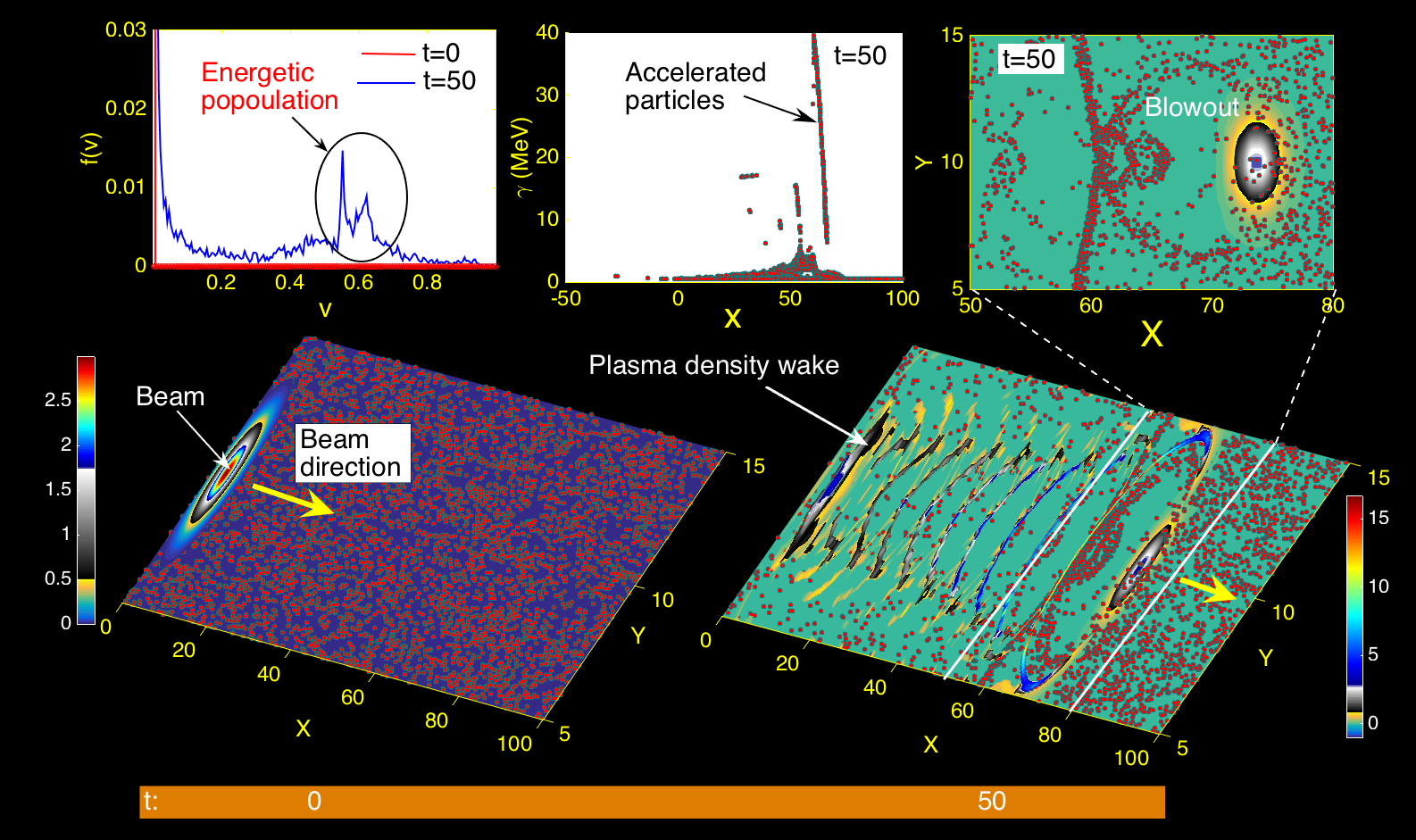}
     \caption{Distribution of the test electrons placed in all over the simulation box at different times ($t=0$ and $t=50$) for a  bi-Gaussian driver beam of energy 3.5 GeV ($v_b=0.99999999$)  and  normalized peak beam density ($n_{b0}$)=2.0, $\sigma_x=\sqrt{2}$ and $\sigma_y=1$ }
    \label{fig10}
\end{figure*}

In this section, we have performed the test particle (electron) simulation to study the energy gain in the process of acceleration. Test electrons are introduced into the fluid simulation and their energy distribution is studied at different times. The dynamics of the test electron are determined by the equation of motion, $\frac{d\vec{p_i}}{dt}=-\vec{E}- (\vec{v_i} \times \vec{B})$; where $p_i=v_i (1-v_i^2)^{-1/2}$ is the  momentum of $i$-th electron having velocity $v_i$. The basic principle for advancing the position and velocity of the test particle in time is based on the Boris pusher algorithm \cite{testboris}. The self-consistent effect of test electrons on the wakefield has been ignored. In our first numerical experiment (shown in (\ref{fig10})), we have randomly distributed 10000 electrons with initial velocity $v_i (t=0)=0$ (extremely cold electrons) in all over the simulation box. The beam ( $n_{b0}=3$, $\sigma_x=\sqrt{2}$, $\sigma_y=1$, $v_b=0.99999999$) is then employed in the simulation which propagates from one end to other end.  The beam creates the wakefield (blowout) which then traps the nearby test electrons in its potential well. The trapped electrons gain energy by the electric field of the blowout and propagate near the speed of beam. We have plotted the velocity distribution function of these particles at $t=0$ and $50$). It is observed that the test electrons can gain maximum 40 MeV energy in a length of 50 ($\frac{c}{\omega_{pe}}$).  
\par
\begin{figure*}
    \includegraphics[width=1.0\textwidth]{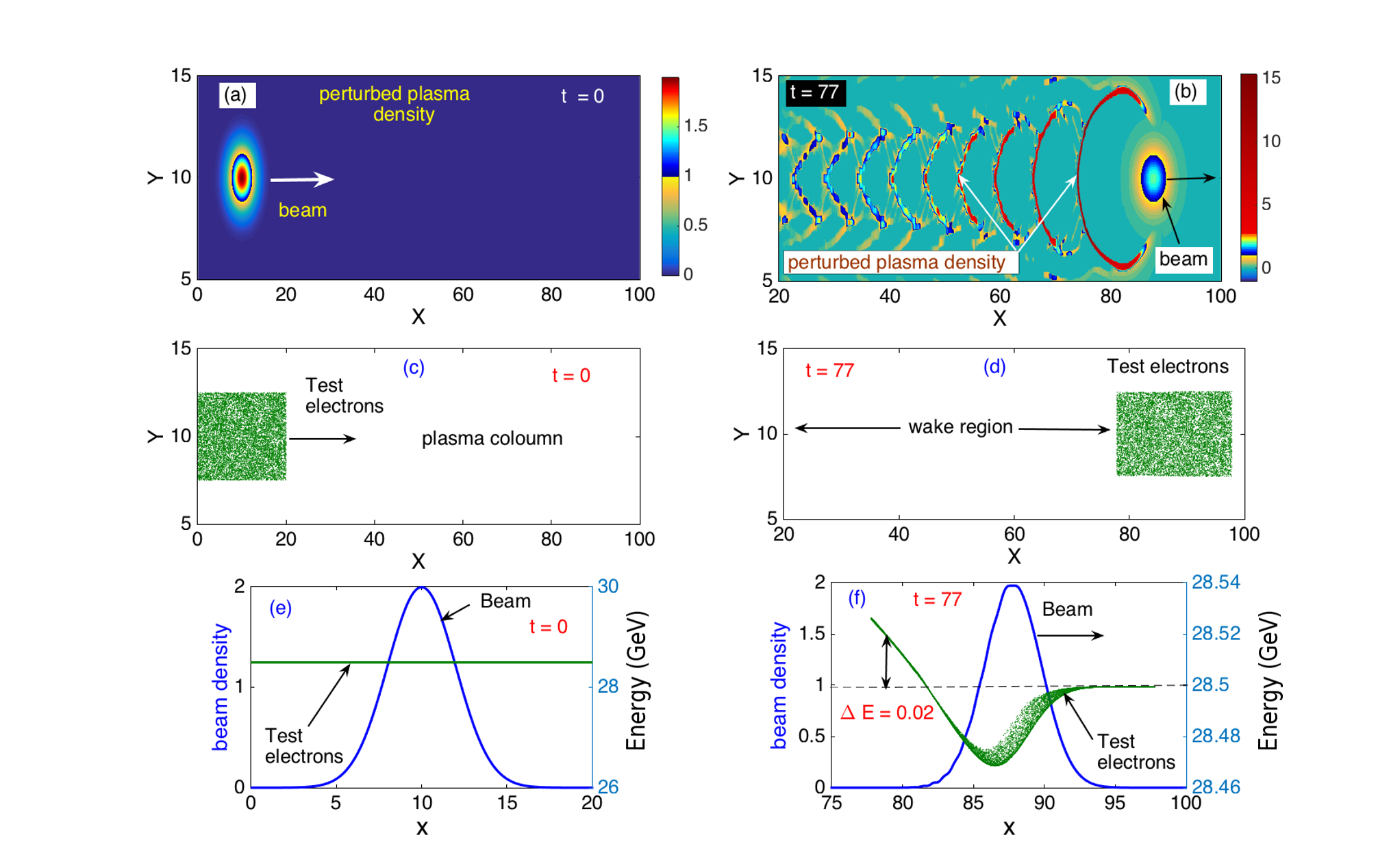}
 \caption{Distribution (space and energy) of the test electrons at different times ($t=0$ and $t=77$) which have been injected at the place of the driver beam of energy 28.5 GeV ($v_b=0.9999999998461$)  and  normalized peak beam density ($n_{b0}$)=2.0, $\sigma_x=2$ and $\sigma_y=1$ }
    \label{fig11}
\end{figure*}

\begin{figure*}
    \includegraphics[width=1.0\textwidth]{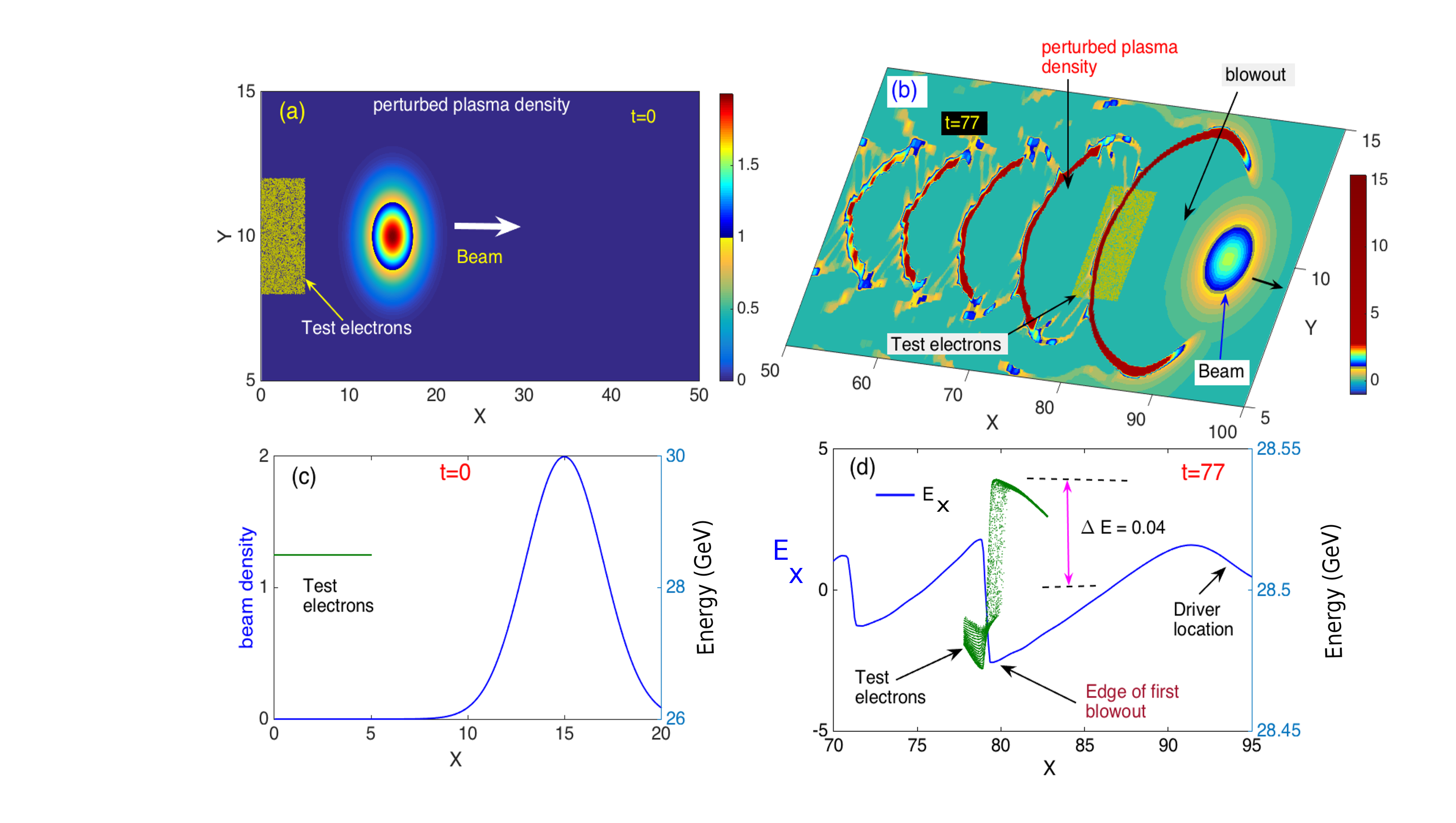}
   \caption{Distribution (space and energy) of the test electrons injected near the axial edge of the first blowout structure at different times ($t=0$ and $t=77$) for the normalized peak beam density ($n_{b0}$)=2.0, $v_b=0.9999999998461$, $\sigma_x=2$ and $\sigma_y=1$ }
    \label{fig12}
\end{figure*}

\vspace{\baselineskip}
Next, we have simulated the experimental results given in ref. \cite{hogan}, where an electron beam having total number of electrons  $N=1.8 \times 10^{10}$, $\sigma_x=20 \mu m$, $\sigma_y= 10 \mu m$ and energy $28.5$ GeV is injected in a plasma ($n_0=2.8 \times 10^{17} cm^{-3}$). Normalizing the quantities, we get,  $n_{b0}=2$, $\sigma_x=2 $, $\sigma_y= 1$,  $v_b=0.9999999998461$. Using these values of the normalized beam parameters and
also with 10000  test electrons as a form of beam having same initial energies of $28.5$ GeV injected with the driver beam, we have performed our simulation  (see Fig. \ref{fig11} at $t=0$). It is observed that the electrons from the front of the test beam lose their energy and the electrons from the back of the beam gain the maximum energy of $200 MeV$ in a length of 77 ($\frac{c}{\omega_{pe}}$) (see Fig. \ref{fig11} at $t=77$). This shows that the electrons can gain maximum energy up to $\sim 2.6$ GeV in a $10$ cm ($ 10^4 \frac{c}{\omega_{pe}}$) long plasma. These results show a good conformity with the experimental results given in ref. \cite{hogan}.  Further, for same beam parameters, we inject the test electron beam near the axial edge of the first blowout structure, where the amplitude of the longitudinal electric field is maximum (shown in fig. (\ref{fig12})). It is observed that the electrons can gain the maximum energy of 400 MeV in a length of 77 ($\frac{c}{\omega_{pe}}$). Therefore, the max. energy gained by these test electrons placed near the blowout can be doubled $\sim 5.2$ GeV after passing 10 cm long plasma.
 
\section{ summary}
The 2-D excitation of relativistic electron beam driven wakefield in a cold plasma is studied using fluid simulation techniques. The simulation results show a good agreement with 1-D results \cite{ratan} for a beam having larger transverse extension compared to the longitudinal extension. It is also shown that, for short and over dense beam, the structure of the excited wake field exhibits blowout which also matches with the analytically modeled results given in ref. \cite{lu_pop}. Further, injecting the test particles in the simulation, we show that the maximum energy gains 2.6 GeV by an electron from the back of the beam of energy 28.5 in a 10 cm long plasma, matches with the earlier experimental observation presented in ref. \cite{hogan}. Using a discrete trailing beam instead of accelerating electrons from the back of the driver, which is placed near the axial edge of the first blowout, the maximum energy gain is found to be doubled $\sim$ 5.2 GeV.
%
%
\bibliographystyle{unsrt}

\begin{thebibliography}{10}

\bibitem{esarey}
E.~Esarey, C.~B. Schroeder, and W.~P. Leemans.
\newblock Physics of laser-driven plasma-based electron accelerators
\newblock {\em Rev. Mod. Phys.}, 81, 1229(2009).


\bibitem{malkasc}
V. Malka, S. Fritzler, E. Lefebvre, M. M. Aleonard, F. Burgy, J. P. Chambaret, J. F. Chemin, K. Krushelnick, G. Malka, S. P. D. Mangles, Z. Najmudin, M. Pittman, J.P. Rousseau, J.N. Scheurer, B. Walton, A. E. Dangor.
\newblock Electron Acceleration by a Wake Field Forced by an Intense Ultrashort Laser Pulse. 
\newblock {\em Science }, 298, 1596 (2002).


\bibitem{cj}
C.~Joshi.
\newblock The development of laser- and beam-driven plasma accelerators as an experimental field.
\newblock {\em Physics of Plasmas}, 14, 055501(2007).

\bibitem{Golovin}
G.~Golovin, S.~Chen, N.~Powers, C.~Liu, S.~Banerjee, J.~Zhang, M.~Zeng, Z.~Sheng and and D.~Umstadter.
\newblock Tunable monoenergetic electron beams from independently controllable laser-wakefield acceleration and injection.
\newblock {\em Phys. Rev. ST Accel. Beams}, 18, 011301(2015).

\bibitem{Faure}
J.~Faure, C.~Rechatin, A.~Norlin, A.~Lifschitz, Y.~Glinec and V.~Malka.
\newblock Controlled injection and acceleration of electrons in plasma wakefields by colliding laser pulses.
\newblock {\em Nature}, 444, 737(2006).

\bibitem{Muggli}
Patric Muggli and Mark J. Hogan.
\newblock Review of high-energy plasma wakefield experiments.
\newblock {\em Comptes Rendus Physique },10, 116(2009).


\bibitem{surf}
Mike Downer and Rafal Zgadzaj.
\newblock Accelerator physics: Surf's up at SLAC.
\newblock {\em Nature}, 515, 40(2014).




\bibitem{chen}
Pisin Chen, J.~M. Dawson, W.~Robert Huff and T.~Katsouleas.
\newblock Acceleration of electrons by the interaction of a bunched electron beam with a plasma.
\newblock {\em Physical review letters}, 54, 693(1985).

\bibitem{uhm}
Han Sup uhm, Glenn Joyce.
\newblock Theory of wake‐field effects of a relativistic electron beam propagating in a plasma.
\newblock {\em Physics of Fluids B: Plasma Physics}, 3, 1587(1991).



\bibitem{katsouleas}
T.~Katsouleas.
\newblock Physical mechanisms in the plasma wake-field accelerator.
\newblock {\em Phys. Rev. A}, 33, 2056(1986).

\bibitem{hogan}
M. J. Hogan, C. D. Barnes, C. E. Clayton, F. J. Decker, S. Deng, P. Emma, C. Huang, R. H. Iverson, D. K. Johnson, C. Joshi, T. Katsouleas, P. Krejcik, W. Lu, K. A. Marsh, W. B. Mori, P. Muggli, C. L. O’Connell, E. Oz, R. H. Siemann, and D. Walz.
\newblock Multi-GeV Energy Gain in a Plasma-Wakefield Accelerator.
\newblock {\em Phys. Rev. Lett..}, 95, 054802  (2005).

\bibitem{barov}
N. Barov, J. B. Rosenzweig, M. E. Conde, W. Gai, and J. G. Power.
\newblock Observation of plasma wakefield acceleration in the underdense regime.
\newblock {\em Phys. Rev. ST Accel. Beams,}, 3, 011301 (2000)


\bibitem{kallos}
Efthymios Kallos, Tom Katsouleas, Wayne D. Kimura, Karl Kusche, Patric Muggli, Igor Pavlishin, Igor Pogorelsky, Daniil Stolyarov, and Vitaly Yakimenko.
\newblock High-Gradient Plasma-Wakefield Acceleration with Two Subpicosecond Electron Bunches.
\newblock {\em Phys. Rev. Lett.,}, 100, 074802  (2008)


\bibitem{Blumenfeld}
Ian Blumenfeld, Christopher E. Clayton, Franz-Josef Decker, Mark J. Hogan, Chengkun Huang,
Rasmus Ischebeck, Richard Iverson, Chandrashekhar Joshi, Thomas Katsouleas, Neil Kirby, Wei Lu, Kenneth A. Marsh, Warren B. Mori, Patric Muggli, Erdem Oz, Robert H. Siemann, Dieter Walz and Miaomiao Zhou.
\newblock Energy doubling of 42 GeV electrons in a metre-scale plasma wakefield accelerator.
\newblock {\em Nature}, 445, 741(2007).


\bibitem{litos}
 M.~Litos, E.~Adli,	W.~An, C.~I. Clarke, C.~E.Clayton, S.~Corde,	J.~P. Delahaye,	R.~J. England, A.~S. Fisher, J.~Frederico, S.~Gessner,	
S.~Z. Green, M.~J. Hogan, C.~Joshi,	
W.~Lu,	K.~A. Marsh, W.~B. Mori,	
P.~Muggli, N.~Vafaei-Najafabadi,	
D.~Walz, G.~White, Z.~Wu, V.~Yakimenko and G.~Yocky.
\newblock High-efficiency acceleration of an electron beam in a plasma wakefield accelerator.
\newblock {\em Nature}, 515, 92(2014).





\bibitem{rosenzweig}
J.~B. Rosenzweig.
\newblock Nonlinear Plasma Dynamics in the Plasma Wake-Field Accelerator.
\newblock {\em Physical Review Letters}, 58, 555(1987).



\bibitem{amatuni}
A.~Ts. Amatuni, S.~Elbakram and E.~V. Sekhpessian.
\newblock {\em Yerevan Physics Institute Report No. ERFI 85-832}, 1985.


\bibitem{ruth}
R.~D.Ruth, A.~Chao, P.~L. Morton and P.~B.Wilson.
\newblock {\em particle Accelerators}, 17, 171(1985).

\bibitem{ratan}
Ratan Kumar Bera, Sudip Sengupta and Amita Das.
\newblock Fluid simulation of relativistic electron beam driven wakefield in a cold plasma .
\newblock {\em Phys. of Plasmas}, 22, 073109 (2015).


\bibitem{katsouleas_2d}
S. Wilks, T. Katsouleas, J. M. Dawson and J. J. Su.
\newblock Beam Loading Efficiency in Plasma Accelerators.
\newblock {\em Pro- ceedings of the 1987 IEEE Particle Accelerator Confer- ence: Accelerator Engineering and Technology}, 1, 100-102 (1987).


\bibitem{chen_2d}
Pisin Chen.
\newblock A possible final focusing mechanism for linear colliders.
\newblock {\em SLAC-PUB-3823 (Rev.)}, SLAC/AP-46, (A/AP), November (1985).


\bibitem{lu_prl}
W. Lu, C. Huang, M. Zhou, W. B. Mori, and T. Katsouleas
\newblock Nonlinear Theory for Relativistic Plasma Wakefields in the Blowout Regime.
\newblock {\em Phys. Rev. Lett.}, 96, 165002 (2006).



\bibitem{osiris}
R A Fonseca, S F Martins, L O Silva, J W Tonge, F S Tsung and W B Mori.
\newblock One-to-one direct modeling of experiments and astrophysical scenarios: pushing the envelope on kinetic plasma simulations.
\newblock {\em Plasma Physics and Controlled Fusion}, 50, 124034 (2008).



\bibitem{epoch}
David Tsiklauri.
\newblock Electron plasma wake field acceleration in solar coronal and chromospheric plasmas.
\newblock {\em Physics of Plasmas},  24, 072902 (2017).

\bibitem{quickpic}
C. Huang, V. K. Decyk, C. Ren, M. Zhou, W. Lu, W. B. Mori, J. H. Cooley, T. M. Antonsen, Jr., and T. Katsouleas.
\newblock Quickpic: A highly efficient particle-in-cell code for modeling wakefield acceleration in plasmas.
\newblock {\em J. Comput. Phys}, 217, 658 (2006).



\bibitem{bulanov}
S. V. Bulanov, F. Pegoraro,  A. M. Pukhov, and A. S. Sakharov.
\newblock Transverse-Wake Wave Breaking
\newblock {\em Phys. Rev. Lett.}, 78, 22 (1997).

\bibitem{lu_pop}
W. Lu , C. Huang,  M. Zhou, M. Tzoufras, F. S. Tsung, W. B. Mori and T. Katsouleas.
\newblock A nonlinear theory for multidimensional relativistic plasma wave wakefields.
\newblock {\em Physics of Plasmas.}, 13, 056709 (2006).
\bibitem{sudip_pre}
Sudip Sengupta, Vikrant Saxena, Predhiman K. Kaw, Abhijit Sen, and Amita Das,
\newblock Phase mixing of relativistically intense waves in a cold homogeneous plasma.
\newblock {\em Phys. Rev. E }, 79, 026404(2009).

\bibitem{prabal}
Prabal Singh Verma, Sudip Sengupta, and Predhiman Kaw
\newblock Breaking of Longitudinal Akhiezer-Polovin Waves.
\newblock {\em Phys. Rev. Lett.}, 108, 125005 (2012).

\bibitem{prabal_pop_2011}
Prabal Singh Vermaa, Sudip Sengupta, and Predhiman K. Kaw.
\newblock Nonlinear evolution of an arbitrary density perturbation in a cold homogeneous unmagnetized plasma
\newblock {\em Physics of Plasmas}, 18, 012301 (2011).

\bibitem{infeld}
E. ~Infeld and G. ~Rowlands,
\newblock Relativistic bursts.
\newblock {\em Phys. Rev. Lett.}, 62, 1122 (1989)

\bibitem{boris}
Jay.~P. Boris, Alexandra M. Landsberg, Elaine S. Oran and John H. Gardner.
\newblock LCPFCT- Flux-corrected Transport Algorithm for Solving Generalized Continuity Equations.
\newblock {\em Naval Research laboratory, Washington}, NRL/MR/6410-93-7192, 1993.


\bibitem{numr}
W.~Press, R.~Assmann, A.~Teukolsky, W.~Vetterling and Brian P. Flannery.
\newblock Numerical Recipes: The Art of Scientific Computing.
\newblock {\em Cambridge University Press}, 1992.


\bibitem{deepa}
Deepa Verma, Ratan Kumar Bera, Amita Das, and Predhiman Kaw.
\newblock The stability of 1-D soliton in transverse direction.
\newblock {\em Physics of Plasmas}, 23, 123102 (2016).


\bibitem{deepa2}
Deepa Verma, Ratan Kumar Bera, Atul Kumar, Bhavesh Patel, and Amita Das.
\newblock Observation of 1-D time dependent non-propagating laser plasma structures using fluid and PIC codes.
\newblock {\em Physics of Plasmas}, 24, 123111 (2017).



\bibitem{testboris}
C. K. Birdsall, and  A.B. Langdon,
\newblock Plasma Physics Via Computer Simulations.
\newblock {\em Institute of Physics Publishing ,} Bristol and Philadelphia, 1991






\end{thebibliography}

%

\newpage
\vspace{10cm}



\end{document}